\documentclass[12pt]{article}
\usepackage{graphics,graphicx}
\usepackage{amssymb,epsfig,amsmath,euscript,array}
\usepackage{cite}
\usepackage{pstricks}
\usepackage{color}

\makeatletter
\@addtoreset{equation}{section}
\makeatother

\newcounter{multieqs}

\newcommand{\be}{\begin{equation}}
\newcommand{\ee}{\end{equation}}

\newcommand{\bm}[1]{\mbox{\boldmath $#1$}}

\newcommand{\kslash}{k \!\!\! / }

\newcommand{\lslash}{l \!\! / }
\newcommand{\Pslash}{P \!\!\!\! / }

\newcommand{\islash}{i \!\!\! / }
\newcommand{\jslash}{j \!\!\! / }
\newcommand{\aslash}{a \!\!\! / }
\newcommand{\bslash}{{b \hspace{-6pt} \slash} }

\newcommand{\onslash}{1 \!\!\! / }
\newcommand{\twslash}{2 \!\!\!/ }
\newcommand{\thslash}{3 \!\!\!/ }
\newcommand{\foslash}{4 \!\!\! / }
\newcommand{\fislash}{5 \!\!\! / }

\newcommand{\mslash}{m \!\!\! / }

\def\bd{\begin{document}}
\def\ed{\end{document}}
\def\nn{\nonumber}
\def\bea{\begin{eqnarray}}
\def\eea{\end{eqnarray}}

\def\red{\color{red}}
\def\black{\color{black}}
\def\blue{\color{blue}}
\def\orange{\color{orange}}

\def\ab{(ijab)}
\def\ba{(ijba)}
\def\ijab{{\tr}_{+}(\islash\, \jslash\, \aslash \, \bslash)}
\def\ijba{{\tr}_{+}(\islash\, \jslash\, \bslash \, \aslash)}
\def\ijaP{{\tr}_{+}(\islash\, \jslash\, \aslash \, \Pslash)}
\def\ijPLa{{\tr}_{+}(\islash\, \jslash\, \Pslash_L \, \aslash)}
\def\ijaPL{{\tr}_{+}(\islash\, \jslash\, \aslash \, \Pslash_L)}
\def\ijPLza{{\tr}_{+}(\islash\, \jslash\, \Pslash_{L;z} \, \aslash)}
\def\ijaPLz{{\tr}_{+}(\islash\, \jslash\, \aslash \, \Pslash_{L;z})}
\def\ijPa{{\tr}_{+}(\islash\, \jslash\, \Pslash \, \aslash)}
\def\iaPb{{\tr}_{+}(\islash\, \aslash\, \Pslash \, \bslash)}
\def\ibPa{{\tr}_{+}(\islash\, \bslash\, \Pslash \, \aslash)}
\def\ijPmu{{\tr}_{+}(\islash\, \jslash\, \Pslash \, \mu)}
\def\ibmuP{{\tr}_{+}(\islash\, \bslash\, \mu \, \Pslash)}
\def\ibmua{{\tr}_{+}(\islash\, \bslash\, \mu \, \aslash)}
\def\iamub{{\tr}_{+}(\islash\, \aslash\, \mu \, \bslash)}
\def\jaPb{{\tr}_{+}(\jslash\, \aslash\, \Pslash \, \bslash)}
\def\ijmuP{{\tr}_{+}(\islash\, \jslash\, \mu \, \Pslash)}
\def\ijmum{{\tr}_{+}(\islash\, \jslash\, \mu \, \mslash)}
\def\ijmmu{{\tr}_{+}(\islash\, \jslash\, \mslash \, \mu)}
\def\ijmP{{\tr}_{+}(\islash\, \jslash\, \mslash \, \Pslash)}
\def\iabP{{\tr}_{+}(\islash\, \aslash\, \bslash \, \Pslash)}
\def\ijbP{{\tr}_{+}(\islash\, \jslash\, \bslash \, \Pslash)}
\def\jbPa{{\tr}_{+}(\jslash\, \bslash\, \Pslash \, \aslash)}
\def\ijPb{{\tr}_{+}(\islash\, \jslash\, \Pslash \, \bslash)}
\def\jbmua{{\tr}_{+}(\jslash\, \bslash\, \mu \, \aslash)}

\def\loablt{ {\tr}_{+}(\lslash_1\, \aslash \, \bslash\, \lslash_2)}

 \def\ijlolt{{\tr}_{+}(\islash\, \jslash\, \lslash_1 \, \lslash_2)}
\def\ijltlo{{\tr}_{+}(\islash\, \jslash\, \lslash_2 \, \lslash_1)}
\def\ibloa{{\tr}_{+}(\islash\, \bslash\, \lslash_1 \, \aslash)}
\def\jaltb{{\tr}_{+}(\jslash\, \aslash\, \lslash_2 \, \bslash)}
\def\ialtb{{\tr}_{+}(\islash\, \aslash\, \lslash_2 \, \bslash)}
\def\bltloa{{\tr}_{+}(\bslash\, \lslash_2\, \lslash_1 \, \aslash)}
\def\jbloa{{\tr}_{+}(\jslash\, \bslash\, \lslash_1 \, \aslash)}
\def\ibPb{{\tr}_{+}(\islash\, \bslash\, \Pslash \, \bslash)}
\def\ijltb{{\tr}_{+}(\islash\, \jslash\, \lslash_2 \, \bslash)}

\def\ijloa{{\tr}_{+}(\islash\, \jslash\,  \lslash_1 \, \aslash)}
\def\ijblt{{\tr}_{+}(\islash\, \jslash\,  \bslash \, \lslash_2)}

\def\jakb{{\tr}_{+}(\jslash\, \aslash\, \kslash \, \bslash)}
\def\iakb{{\tr}_{+}(\islash\, \aslash\, \kslash \, \bslash)}

\def\tofo{{\tr}_{+}(\onslash\, \thslash\, \twslash \, \foslash)}
\def\foto{{\tr}_{+}(\onslash\, \thslash\, \foslash \, \twslash)}
\def\tofi{{\tr}_{+}(\onslash\, \thslash\, \twslash \, \fislash)}
\def\fito{{\tr}_{+}(\onslash\, \thslash\, \fislash \, \twslash)}

\def\lrangle#1#2{\langle #1\,#2\rangle}

\def\Li{{\rm Li}}
\def\eps{\epsilon}
\def\epsuv{{\epsilon_{\rm \mbox{\tiny UV}}}}
\let\bm=\bibitem
\let\la=\label

\def\npb#1#2#3{Nucl. Phys. {\bf{B#1}} #3 (#2)}
\def\plb#1#2#3{Phys. Lett. {\bf{#1B}} #3 (#2)}
\def\prl#1#2#3{Phys. Rev. Lett. {\bf{#1}} #3 (#2)}
\def\prd#1#2#3{Phys. Rev. {D \bf{#1}} #3 (#2)}
\def\cmp#1#2#3{Comm. Math. Phys. {\bf{#1}} #3 (#2)}
\def\cqg#1#2#3{Class. Quantum Grav. {\bf{#1}} #3 (#2)}
\def\nppsa#1#2#3{Nucl. Phys. B (Proc. Suppl.) {\bf{#1A}}#3 (#2)}
\def\ap#1#2#3{Ann. of Phys. {\bf{#1}} #3 (#2)}
\def\ijmp#1#2#3{Int. J. Mod. Phys. {\bf{A#1}} #3 (#2)}
\def\rmp#1#2#3{Rev. Mod. Phys. {\bf{#1}} #3 (#2)}
\def\mpla#1#2#3{Mod. Phys. Lett. {\bf A#1} #3 (#2)}
\def\jhep#1#2#3{J. High Energy Phys. {\bf #1} #3 (#2)}
\def\atmp#1#2#3{Adv. Theor. Math. Phys. {\bf #1} #3 (#2)}
\newcommand{\EQ}[1]{\begin{equation} #1 \end{equation}}
\newcommand{\AL}[1]{\begin{subequations}\begin{align} #1 \end{align}\end{subequations}}
\newcommand{\SP}[1]{\begin{equation}\begin{split} #1 \end{split}\end{equation}}
\newcommand{\ALAT}[2]{\begin{subequations}\begin{alignat}{#1} #2 \end{alignat}
                        \end{subequations}}
\def\beqa{\begin{eqnarray}}
\def\eeqa{\end{eqnarray}}
\def\beq{\begin{equation}}
\def\eeq{\end{equation}}
\def\sst{\scriptscriptstyle}
\def\thetabar{\bar\theta}
\def\Tr{{\rm Tr}}
\def\one{\mbox{1 \kern-.59em {\rm l}}}
 \def\Nh{\hat{N}}

\newcommand{\half}{{\textstyle {1 \over 2}}}

\def\a{\alpha}      \def\da{{\dot\alpha}}
\def\b{\beta}       \def\db{{\dot\beta}}
\def\c{\gamma}  \def\G{\Gamma}  \def\cdt{\dot\gamma}
\def\d{\delta}  \def\D{\Delta}  \def\ddt{\dot\delta}
\def\e{\epsilon}        \def\vare{\varepsilon}
\def\f{\phi}    \def\F{\Phi}    \def\vvf{\f}
\def\h{\eta}
\def\k{\kappa}
\def\l{\lambda} \def\L{\Lambda}
\def\m{\mu} \def\n{\nu}
\def\o{\omega}
\def\p{\pi} \def\P{\Pi}
\def\r{\rho}
\def\s{\sigma}  \def\S{\Sigma}
\def\t{\tau}
\def\th{\theta} \def\Th{\Theta} \def\vth{\vartheta}
\def\X{\Xeta}
\def\z{\zeta}
\def\de{\partial}

\def\cA{{\cal A}} \def\cB{{\cal B}} \def\cC{{\cal C}}
\def\cD{{\cal D}} \def\cE{{\cal E}} \def\cF{{\cal F}}
\def\cG{{\cal G}} \def\cH{{\cal H}} \def\cI{{\cal I}}
\def\cJ{{\cal J}} \def\cK{{\cal K}} \def\cL{{\cal L}}
\def\cM{{\cal M}} \def\cN{{\cal N}} \def\cO{{\cal O}}
\def\cP{{\cal P}} \def\cQ{{\cal Q}} \def\cR{{\cal R}}
\def\cS{{\cal S}} \def\cT{{\cal T}} \def\cU{{\cal U}}
\def\cV{{\cal V}} \def\cW{{\cal W}} \def\cX{{\cal X}}
\def\cY{{\cal Y}} \def\cZ{{\cal Z}}

\def\ua{\underline{\alpha}}
\def\ub{\underline{\phantom{\alpha}}\!\!\!\beta}
\def\uc{\underline{\phantom{\alpha}}\!\!\!\gamma}
\def\um{\underline{\mu}}
\def\ud{\underline\delta}
\def\ue{\underline\epsilon}
\def\una{\underline a}\def\unA{\underline A}
\def\unb{\underline b}\def\unB{\underline B}
\def\unc{\underline c}\def\unC{\underline C}
\def\und{\underline d}\def\unD{\underline D}
\def\une{\underline e}\def\unE{\underline E}
\def\unf{\underline{\phantom{e}}\!\!\!\! f}\def\unF{\underline F}
\def\unm{\underline m}\def\unM{\underline M}
\def\unn{\underline n}\def\unN{\underline N}
\def\unp{\underline{\phantom{a}}\!\!\! p}\def\unP{\underline P}
\def\unq{\underline{\phantom{a}}\!\!\! q}
\def\unQ{\underline{\phantom{A}}\!\!\!\! Q}
\def\unH{\underline{H}}

\def\As {{A \hspace{-6.4pt} \slash}\;}
\def\bs {{b \hspace{-6.4pt} \slash}\;}
\def\Ds {{D \hspace{-6.4pt} \slash}\;}
\def\ds {{\del \hspace{-6.4pt} \slash}\;}
\def\ss {{\s \hspace{-6.4pt} \slash}\;}
\def\ks {{ k \hspace{-6.4pt} \slash}\;}
\def\ps {{p \hspace{-6.4pt} \slash}\;}
\def\pas {{{p_1} \hspace{-6.4pt} \slash}\;}
\def\pbs {{{p_2} \hspace{-6.4pt} \slash}\;}
\def\Ps {{P \hspace{-6.4pt} \slash}\;}
\def\Qs {{Q \hspace{-6.4pt} \slash}\;}

\def\Fh{\hat{F}}
\def\Vh{\hat{V}}
\def\Xh{\hat{X}}
\def\ah{\hat{a}}
\def\xh{\hat{x}}
\def\yh{\hat{y}}
\def\ph{\hat{p}}
\def\xih{\hat{\xi}}
\def\psit{\tilde{\psi}}
\def\Psit{\tilde{\Psi}}
\def\tht{\tilde{\th}}
\def\lt{\tilde{\lambda}}
\def\hl{\hat{\lambda}}
\def\hlt{\hat{\tilde{\lambda}}}
\def\llt{\tilde{l}}
\def\At{\tilde{A}}
\def\Qt{\tilde{Q}}
\def\Rt{\tilde{R}}
\def\Nt{\tilde{N}}

\def\at{\tilde{a}}
\def\st{\tilde{s}}
\def\ft{\tilde{f}}
\def\pt{\tilde{p}}
\def\qt{\tilde{q}}
\def\vt{\tilde{v}}
\def\nt{\tilde{n}}

\def\delb{\bar{\partial}}
\def\bz{\bar{z}}
\def\bD{\bar{D}}
\def\bB{\bar{B}}

\def\bk{{\bf k}}
\def\bl{{\bf l}}
\def\bp{{\bf p}}
\def\bq{{\bf q}}
\def\br{{\bf r}}
\def\bx{{\bf x}}
\def\by{{\bf y}}
\def\bR{{\bf R}}
\def\bV{{\bf V}}

\def\d{\delta}\def\D{\Delta}\def\ddt{\dot\delta}
\def\pa{\partial} \def\del{\partial}
\def\xx{\times}
\def\uno{\mbox{1 \kern-.59em {\rm l}}}
\def\trp{^{\top}}
\def\inv{^{-1}}
\def\dag{{^{\dagger}}}
\def\pr{^{\prime}}
\def\lan{\langle}
\def\ran{\rangle}
\def\rar{\rightarrow}
\def\lar{\leftarrow}
\def\lrar{\leftrightarrow}
\newcommand{\0}{\,\!}      
\def\one{1\!\!1\,\,}
\def\im{\imath}
\def\jm{\jmath}
\newcommand{\tr}{\mbox{tr}}
\newcommand{\slsh}[1]{/ \!\!\!\! #1}
\def\vac{|0\rangle}
\def\lvac{\langle 0|}
\def\hlf{\frac{1}{2}}
\def\ove#1{\frac{1}{#1}}
\def\Box{\square}
\def\ZZ{\mathbb{Z}}
\def\CC#1{({\bf #1})}
\def\bcomment#1{}
\def\bfhat#1{{\bf \hat{#1}}}
\def\VEV#1{\left\langle #1\right\rangle}
\newcommand{\ex}[1]{{\rm e}^{#1}} \def\ii{{\rm i}}
\def\rr{{\rm r}} \def\rs{{\rm s}}\def\rv{{\rm v}}
\def\ri{{\rm i}}\def\rj{{\rm j}}
\newcommand{\lrbrk}[1]{\left(#1\right)}
\newcommand{\sfrac}[2]{{\textstyle\frac{#1}{#2}}}

\font\mybb=msbm10 at 12pt
\def\bb#1{\hbox{\mybb#1}}

\font\myBB=msbm10 at 18pt
\def\BB#1{\hbox{\myBB#1}}

\begin{document}

\begin{flushright}
IPPP/10/27,
DCPT/10/54
\end{flushright}

\vspace{3pt}

\begin{center}

\hspace{-0.8cm}{\Large \bf
Regular Wilson loops and MHV amplitudes \\  at weak and strong coupling
}

\vspace{11pt}
\end{center}

\centerline{\mbox {\large Paul~Heslop$^{a}$ and
Valentin~V.~Khoze$^{b}$}%
{
\renewcommand{\thefootnote}{}  \footnotetext{
{\tt \{paul.heslop, valya.khoze\}@durham.ac.uk}
}
}
}

\begin{center}
{\small \em
\begin{itemize}
\item[\ \ \ \ \ \ $^a$]
Institute for Particle Physics Phenomenology,  \\
Department of Mathematical Sciences and Department of Physics\\
Durham University,
Durham, DH1 3LE, United Kingdom\\
\item[\ \ \ \ \ \ $^b$]
Institute for Particle Physics Phenomenology,  \\
Department of Physics,
Durham University, \\
Durham, DH1 3LE, United Kingdom

\end{itemize}
}

\vspace{-8pt}

\vspace{23pt} {\bf Abstract}
\end{center}

\noindent
Traditionally,  the duality between Wilson loops and amplitudes beyond
one loop in
$\cN$=4 SYM is characterised by the remainder function. Because of the
perturbative origins of the BDS expression, the remainder function is
more natural at weak than at strong coupling.
We advocate instead a more direct  approach,   based on considering
ratios of Wilson loops. This allows us to define a manifestly finite,
regularisation independent, conformally invariant quantity. It does
not make a direct reference to the BDS expression and the definition
is regularisation independent. It is a  natural
object at weak and at
strong coupling, and in the latter case is directly related to the
free energy of an auxiliary integrable system. We then compute
these ratios for continuous families of regular polygons for $6,8$ and
$10$ points at one and two-loops. These results are compared to
expressions derived recently at strong coupling.

\setcounter{page}{0}
\thispagestyle{empty}
\newpage


\section{Introduction}
\label{sec:1}
\setcounter{footnote}{0}

It has been conjectured~\cite{am,dks,bht} that in planar $\cal N$=4 super
Yang-Mills (SYM) there is a non-trivial
relation between scattering amplitudes and Wilson loops,
\beq
\label{wil}
W_n \, :=\, W[ \cC_n]  \ = \ {\rm Tr} \, \cP \exp \left[ i g\oint_{\cC_n} \! d\tau  \ \dot{x}^{\mu} (\tau )A_\mu (x(\tau ))   \right]
\ ,
\eeq
with a lightlike $n$-edged polygonal contour $\cC_n$ obtained by  attaching the momenta of the scattered particles
$p_1, \ldots , p_n$ one after the other, following the order
of the colour generators in the colour-ordered scattering amplitude. The vertices, $x_i$, of the polygon are related to the
external momenta via $p_i= x_i-x_{i+1},$ where $x_{n+1}=x_1$. There has been an increasing amount of evidence in support of this amplitude/Wilson loop duality relation~\cite{am,dks,bht,dhks4,dhks5,seven,dhks6,Anastasiou:2009kn}.

For MHV amplitudes, $\cA_n^{\rm MHV}= \cA_n^{\rm MHV\, tree} \times
\cM_n,$, the MHV amplitude/Wilson loop duality~\cite{am,dks,bht} at one loop states simply that
$\cM_n^{(1)}=W_n^{(1)}+{\rm{const}}$~\cite{bht},  whereas beyond one loop
it is normally understood in terms of the
remainder function. The remainder function of the amplitude
${\cR}_n$ (or of the Wilson loop ${\cR}^{WL}_n$) is defined as the
difference between the logarithm of the entire amplitude $\cM_n$
(Wilson loop $W_n$) and the known BDS expression obtained
in~\cite{abdk,bds}, so that
\begin{align}\label{eq:1}
  {\cR}_n&=\log (\cM_n) -(BDS)_n \nonumber\\
  {\cR}^{WL}_n&=\log (W_n) -(BDS)^{WL}_n\ .
\end{align}
The BDS expressions for both the amplitude and the Wilson loop can be found
in the Appendix where we also outline the difference between  the amplitude
and Wilson loop expressions.
The duality then states that
the two remainder functions are identical~\cite{seven,dhks6,Anastasiou:2009kn} (and in particular no constant shifts are allowed)
\begin{equation}\label{r=r}
   {\cR}_n=  {\cR}^{WL}_n\ .
\end{equation}

The amplitudes and Wilson loops are themselves divergent quantities. The amplitudes contain infrared and the Wilson loops
ultraviolet divergences. These divergences break the (dual) conformal
symmetry of the theory.
The remainder function however is a quantity which is constructed to be manifestly
finite since the divergences are cancelled by the BDS contributions.
Furthermore, the remainder functions are known to be conformally
invariant   and as such they depend on the kinematics only through the
conformal cross-ratios $u_{ij}$.~\footnote{More precisely the Wilson loop remainder is
  invariant under conformal transformations in $\cN$=4 super Yang-Mills~\cite{dks}.
  Assuming the Wilson loop/amplitude duality~(\ref{r=r}), the amplitude remainder
  function must then also only depend on conformal cross-ratios. This
  has become known as the dual conformal invariance of the amplitude.}

The BDS expressions are essentially determined by the one loop amplitudes and
as such the first non-vanishing contribution to the remainder
function appears at two loops. Even though BDS expressions have their origin
at one loop in perturbation theory, they do depend on the coupling $a$ through cusp anomalous dimensions
(as recalled in the Appendix) and are straightforwardly extended to all
values of the coupling. Thus the remainder functions in~(\ref{r=r})
are defined for all values of the coupling, and in particular can be
constructed at strong or at weak coupling. Traditionally
computations for Wilson loops or amplitudes beyond one loop have always
been interpreted in terms of the remainder function at weak  and at strong
coupling.

However, the one-loop perturbative origin of the BDS expression means
that the remainder function is not the most natural quantity
appearing at strong coupling. We would like to formulate an
approach which does not involve $\cR_n$ and that allows for a more direct comparison of weak and strong
coupling results. In the following section we will define such a
quantity in terms of a ratio of a Wilson loop with an appropriately
defined reference Wilson loop. We will further argue that this object
is conformally invariant and provides a natural formulation of the
Wilson loop/amplitude duality.

We will then compute
this finite conformal ratio for
hexagons, octagons and decagons. For concreteness
and to keep the kinematics manageable we will concentrate on families of
regular polygons. Strong coupling computations for these regular
polygons  were computed very recently in~\cite{amsv}.
We will compare and supplement these strong coupling answers with
one-loop and two-loop results.

In section~\ref{reg}  we define the regular kinematics, in section~\ref{6pnt}
we present the results for the hexagons, sections~\ref{8pnt} and~\ref{10pnt} contain our
analysis of regular octagons and decagons.

\section{The finite conformally invariant ratio}
\label{sec:2}

At strong coupling, $a\rightarrow \infty$,  the quantity of interest
is $\sqrt{2a} A$ where $A$ is the area of a world sheet ending on
the polygonal Wilson loop\cite{am}. The area is infinite (which is a
reflection of the divergences of the amplitude/Wilson loop) and
needs to be regularised. However we do not wish to rely on any
specific scheme (for example dimensional regularisation which is the
standard choice made in weak coupling computations is not natural at
strong coupling and
not what is used in practice there~\cite{am8,agm,amsv,Hatsuda:2010cc}). Thus we will
construct a manifestly finite quantity, which should be independent of the
regularisation used.

In general the area is a Lorentz invariant quantity which depends on
two-particle invariants, $s_i=(p_i+p_{i+1})^2$, and multi-particle invariants,
$t^{[r>2]}_i=(p_i+p_{i+1} + \dots + p_{i+r-1})^2$.
It can be represented as~\cite{am8}
\begin{equation}\label{Acutfin}
  A(s,t)=A_{\mathrm{cutoff}}(s) + A_{\mathrm{finite}}(s,t)\, , \qquad
  A_{\mathrm{cutoff}}(s)=  4 \int_{\Sigma_0\, ,\,
    z_{\mathrm{AdS}}>\epsilon_c} d^2 w\ .
\end{equation}
Here $\Sigma_0$ is an
appropriate simplified surface in  $AdS_5$ and $\epsilon_c$ is the
cutoff in the radial direction. The important point for us is that
$A_{\mathrm{cutoff}}$ depends on the kinematics only through the
two-particle invariants $s_i$.
In particular~\cite{am8,agm,amsv},
\begin{equation}\label{Acuteven}
A_{\mathrm{cutoff}} = \frac{1}{8} \sum_{i=1}^{n} \,\left(\log \epsilon_c^2 s_i \right)^2\,
-\, \frac{1}{8} \sum_{i=1}^{n} \,\left( \left(\log  s_i \right)^2 +
\sum_{k=0}^{(n-2)/4} (-1)^{k+1}\log  s_i \log  s_{i+1+2k} \right)
\end{equation}
for $n$ even, and a very similar formula holds for odd values of $n$,
\begin{equation}\label{Acutodd}
A_{\mathrm{cutoff}} = \frac{1}{8} \sum_{i=1}^{n} \,\left(\log \epsilon_c^2 s_i \right)^2\,
-\, \frac{1}{4} \sum_{i=1}^{n} \,\left( \left(\log  s_i \right)^2 +
\sum_{k=0}^{2K} (-1)^{k+1}\log  s_i \log  s_{i+1+2k} \right)
\end{equation}
where $n=4K+1$ or $n=4K+3$.

We can thus consider the difference
between the area $A(s,t)$ and a reference area $A(s, \tilde{t})$ which
has the same values of two-particle invariants, whilst the
multi-particle invariants $\tilde{t}$ are fixed. The divergent contributions,
$A_{\mathrm{cutoff}}$, cancel and the difference
between the two areas is manifestly
finite.

This cancellation of divergences in the difference between the area- and the reference area-like quantities
is a general feature which holds at strong and at weak coupling to all orders in perturbation theory, both for the
scattering amplitudes and the Wilson loops. In the context of Wilson loops, the difference of areas is simply
the logarithm of the ratio between the two Wilson loops,
\begin{align}\label{DWn}
  \log \left({W_n \over W_n^{\rm ref}}\right)\  :=\  w_n (s,t) -w_n^{\rm
    ref}(s,\tilde{t})\ = \ {\rm finite} \ .
\end{align}
This is a finite quantity at weak coupling as well, since all the
divergences of each $w_n$, in dimensional regularisation $D=4-2\epsilon$,  are
of the form (see Appendix)
\begin{align}\label{divvvv}
 - {1\over 2 \e^2} \ \sum_{L=1}^{\infty} a^L \, {f^{(L)} (\epsilon)
   \over L^2} \,  \sum_{i=1}^n   \left( - {s_i}
    \over \mu_\mathrm{}^2 \right)^{-L\e} \ ,
\end{align}
ie they depend only on two-particle invariants and thus cancel between the two terms on the right hand side of \eqref{DWn}.
Exactly the same argument holds in the amplitudes case for $\log
({\cM_n / \cM_n^{\rm ref}}).$ Indeed in any regularisation the
divergences will depend only on two-particle invariants\footnote{All infrared divergences of colour-ordered
scattering amplitudes at large $N$ arise from the Sudakov form-factors associated to each consecutive pair of external legs.}, and hence
this argument is independent of any specific regularisation scheme.

To proceed, we note that the number of independent multi-particle invariants
$t_i^{[r]}$ is $n(n-5)/2$, precisely the same as the number of independent cross-ratios
$u_{ij}$ defined as
\begin{equation}\label{uijdef}
  u_{ij}={x_{ij+1}^2 x^2_{i+1 j} \over x_{ij}^2 x_{i+1 j+1}^2} \ ,
\end{equation}
where $x_i$ are the vertices of the polygonal contour of the Wilson loop.
This matching holds when we do not impose the Gramm
determinant constraints, ie the external momenta are not restricted to
four dimensions~\cite{Anastasiou:2009kn}~\footnote{Conformal
  invariance of the remainder function indeed did not require imposing
the Gramm determinant
constraint~\cite{seven,dhks6,Anastasiou:2009kn}.}. Furthermore, as
long as the number of external particles (or for the Wilson
loop the number of edges) $n$ is not divisible by four we can trade
all multi-particle invariants for the cross-ratios $\{s,t\}
\rightarrow \{s,u\}$. Thus the area can
be recast as a function of $s_i$ and $u_{ij}$.

In the special case when $n=4K$, some of the cross-ratios depend only
on two-particle invariants and a complete separation between the $s$-variables and $u$-variables is not
possible. This case will be treated separately in section~\ref{8pnt} where we will show that special kinematical
subsets
for $n=4K$ can still be treated in essentially the same way.
Until then we will concentrate
on cases with $n$ not divisible by four and adopt the $ \{s,u\}$ basis.

The next point to note is that in the $ \{s,u\}$ basis the second term in the first equation in \eqref{Acutfin} does not depend on
the two-particle invariants, $A_{\mathrm{finite}}= A_{\mathrm{finite}}(u)$. This is a consequence of the fact~\cite{am8}
that the second term on the right hand side of \eqref{Acuteven} (or \eqref{Acutodd} for odd values of $n$) is a solution
of the anomalous Ward identities for broken conformal invariance\footnote{In this sense this contribution is the strong-coupling analogue
of the BDS expression which is also a particular solution of the anomaly equations; it was termed `BDS-like' in \cite{am8}.
For $n=4,5$ the BDS and the BDS-like expressions are identical, but for general $n \ge 6$ they are different, in particular
BDS depends on multi-particle invariants while BDS-like does not. For $n \neq 4K$ BDS-like is the unique solution of this
anomaly equation which depends on the kinematics only through two-particle invariants.}
\cite{dks}. Since the entire area (or more precisely the
finite area which excludes the explicit cut-off $\epsilon_c$-dependence) must also satisfy the same conformal anomaly equation,
any finite correction to the second term in \eqref{Acuteven} must be a solution to the corresponding {\it homogeneous} equation.
$A_{\mathrm{finite}}$ is precisely the finite part of the area missed by the cut-off contribution, it must be a solution to the
homogeneous equation, and as such is a function solely of the cross-ratios $u_{ij}$.
In Refs.~\cite{am8,agm,amsv} $A_{\mathrm{finite}}(u)$ was itself
represented as the sum of a few terms,
$A_{\mathrm{finite}}= A_{\mathrm{free}}+A_{\mathrm{periods}}+A_{\mathrm{extra}}$ each of which was derived from independent
reasoning and with particular attention payed to the free energy contribution determined through the
Thermodynamic Bethe Ansatz (TBA) equations.
At present it is not known if and how the TBA structure can arise in the opposite weak coupling limit, thus
in this paper we shall not attempt to break $A_{\mathrm{finite}}$ into separate master functions at weak coupling and
instead concentrate on the $u$-dependence of the entire area.
It is worthwhile pointing out however, that for the $Z_n$-symmetric Wilson loops (which are the main computational application
in this paper), the strong coupling limit of
$\log (W_n / W_n^{\rm ref})$ is equivalent to the free-energy contribution, $A_{\mathrm{free}}$.
The remaining terms, $A_{\mathrm{periods}}$ and $A_{\mathrm{extra}}$, will not contribute to this ratio in the $Z_n$-symmetric case.
In other words, the quantity $\log (W_n / W_n^{\rm ref})$ will allow one to zoom in on the analogue of the the free-energy contribution
at weak coupling.

It follows that the difference between the area and the reference area,
\begin{equation}\label{Adiff}
  A(s,u) - A(s,\tilde{u}) \, =\,  A_{\mathrm{finite}}(u) - A_{\mathrm{finite}}(\tilde{u}) \,\equiv\, f(u,\tilde{u})
\end{equation}
is a finite and conformally-invariant function in the sense that it
depends on the kinematics only through the conformal cross-ratios
$u_{ij}$ as well as the reference ratios $\tilde{u}_{ij}$ which we view as fixed. Manifest finiteness of the area difference
also implies that it can be computed in any regularisation scheme.

Note that exactly the same reasoning applies, in general,  not only to the strong-coupling regime, but also
at weak coupling to all orders in perturbation theory and both for
scattering amplitudes and Wilson loops.
For Wilson loops, as we already discussed, the relevant quantity is
the logarithm of the ratio between the two Wilson loops,
\begin{align}\label{DWnagain}
  \log \left({W_n \over W_n^{\rm ref}}\right)\  :=\  w_n (s,u) -w_n^{\rm
    ref}(s,\tilde{u})\ = \ {\rm finite} (u,\tilde{u})\ .
\end{align}
Once again the right-hand side is a function of conformal cross-ratios
only. This is ensured by the fact that $w_n(s,u)$ satisfies anomalous
conformal Ward identities a particular solution of which
is~(\ref{Acuteven}),(\ref{Acutodd}) which crucially depends only on
two-particle invariants. Thus the difference of two $w$'s with the
same two-particle invariants satisfies non-anomalous conformal Ward
identities and as such is a function of cross-ratios only.

The Wilson loop/MHV amplitude duality then takes a remarkably simple
form in terms of the ratio. It is the statement that
\begin{align}\label{DWMHVA}
 \log \left({W_n \over W_n^{\rm ref}}\right)\, (u,\tilde{u})\  =\ \log \left({\cM_n \over \cM_n^{\rm ref}}\right)\, (u,\tilde{u})
 \ ,
\end{align}
or even more simply
\begin{align} \label{DWMHVAa}
\boxed{  \quad {W_n \over W_n^{\rm ref}}\  =\ {\cM_n \over \cM_n^{\rm
      ref}}\ .\quad }
\end{align}
where we treat the reference variables $\tilde{u}$ as fixed.
This duality correspondence is in terms of manifestly finite,
conformally-invariant quantities. It is defined in a
regularisation-independent manner\footnote{For example one can compute the
  ratio $\cM_n / \cM_n^{\rm
      ref}$ on the Coulomb phase of $\cN$=4 SYM where infrared
  divergences are regulated by the masses as explained in~\cite{Alday:2009zm}.})
and is expected to hold at {\em all}  values of the coupling
constant including, of course, one loop where it is a non-trivial
statement (unlike the formulation in terms of $\cR$).

We would like to add a word of caution in respect to what is meant by the conformal invariance of
\eqref{DWnagain}-\eqref{DWMHVAa}. The right hand side of equation \eqref{DWnagain} is conformally invariant
{\it after} all the dependence on the two-particle invariants is cancelled between $W_n$ and $W_n^{\rm ref}$.
However, the actual ratio was constructed in a given frame where we chose the $s$-variables in
$W_n^{\rm ref}(s,\tilde{u})$ to be the same as in the original Wilson loop $W_n(s,u)$. If one performed a conformal
transformation on the individual Wilson loops before computing their ratio, the $s$-variables would no longer match.

Another interesting observation is that we were free to remove the log
in~(\ref{DWMHVA}) without any loss of information because the ratio is
finite.  This should be compared with the amplitude/Wilson loop itself
and its log (keeping, in both
cases, only  terms up to $O(\epsilon^0)$ ) which contain different pieces of
information.
The reason for this is that
$1/\epsilon^2$ contributions in the singular part of the exponent can hit
$O(\epsilon)$ pieces resulting in $1/\epsilon$ contributions. These
can not be seen from the log of the amplitude itself. For example the five-point two loop amplitude contains parity
odd pieces at $O(1/\epsilon)$ whereas in the log of the amplitude they cancel and appear
only at $O(\epsilon)$~\cite{2l5pt}\footnote{Subsequently this has
  also been seen in the high energy regime in~\cite{5ptregge}.}. The ratio of the
amplitudes is however free from such effects.
Interestingly, it was pointed out in~\cite{Alday:2009zm} that
amplitudes in the
Coulomb branch regularisation are also free from such effects.
Our ratio is of course defined independently of the regularisation in
the first place.

For generic non-MHV amplitudes in $\cN=4$ SYM the quantity
\begin{align}
{  \cA_n^{\rm{NMHV}}(s,t,h)\over \cA_n^{\rm{ref}}(s,\tilde t,\tilde h)}
\end{align}
is a finite quantity~\footnote{Indeed this quantity should be infrared
  finite in any large $N$ gauge theory.}.
 Here $h,\tilde h$ denote helicities and particle
types. Indeed, it has a nice interpretation as the
ratio of hard amplitudes. Infrared factorisation arguments~\cite{Mueller:1979ih,Collins:1980ih,Sen:1981sd,Magnea:1990zb},
dictate that any amplitude is a  product of soft, jet and hard
amplitudes, where the hard amplitude is an infrared safe quantity.
In planar perturbation theory (where we work) the jet times soft
amplitude becomes simply a product of (square roots of) Sudakov form-factors
and as such they depend only on two-particle invariants. What
survives in the ratio is the ratio of hard amplitudes.

If we factor out the corresponding tree level amplitude, then the
object
\begin{align}
{  \left(\cA_n^{\rm{NonMHV}}/  \cA_n^{\rm{NonMHV\, tree}} \right)(s,u,h)
\over
 \left( \cA_n^{\rm{ref}}/  \cA_n^{\rm{ref\,tree}}\right)(s,\tilde u,
 \tilde h)}\ .
\end{align}
is not only finite but we expect it to be (dual)
conformally invariant in view of the expected dual conformal
properties of non-MHV
amplitudes~\cite{Drummond:2008vq,Brandhuber:2008pf,Brandhuber:2009kh}.
Normalisation of all amplitudes by the tree-level factor
also naturally fits with the proposal~\cite{Abel:2007mw} that in the
strong coupling regime any generic
(MHV or NonMHV) amplitude is the product of the corresponding tree-level
amplitude and the helicity-independent
exponential factor involving the same semiclassical action (area) as the
one found for MHV amplitudes in \cite{am}.
Of course the Wilson loop dual for non-MHV amplitudes is not known at
present.

In the rest of the paper we will concentrate on the ratio of Wilson
loops~(\ref{DWnagain}) in $\cN=4$ SYM at large $N$.
In the next section we will define the kinematics for the continuous families of
regular polygons which will be needed for our applications.

\vskip1cm

\centerline{A Note on multi-collinear limits}

\begin{figure}[h!]
  \centering
  \includegraphics[width=13cm]{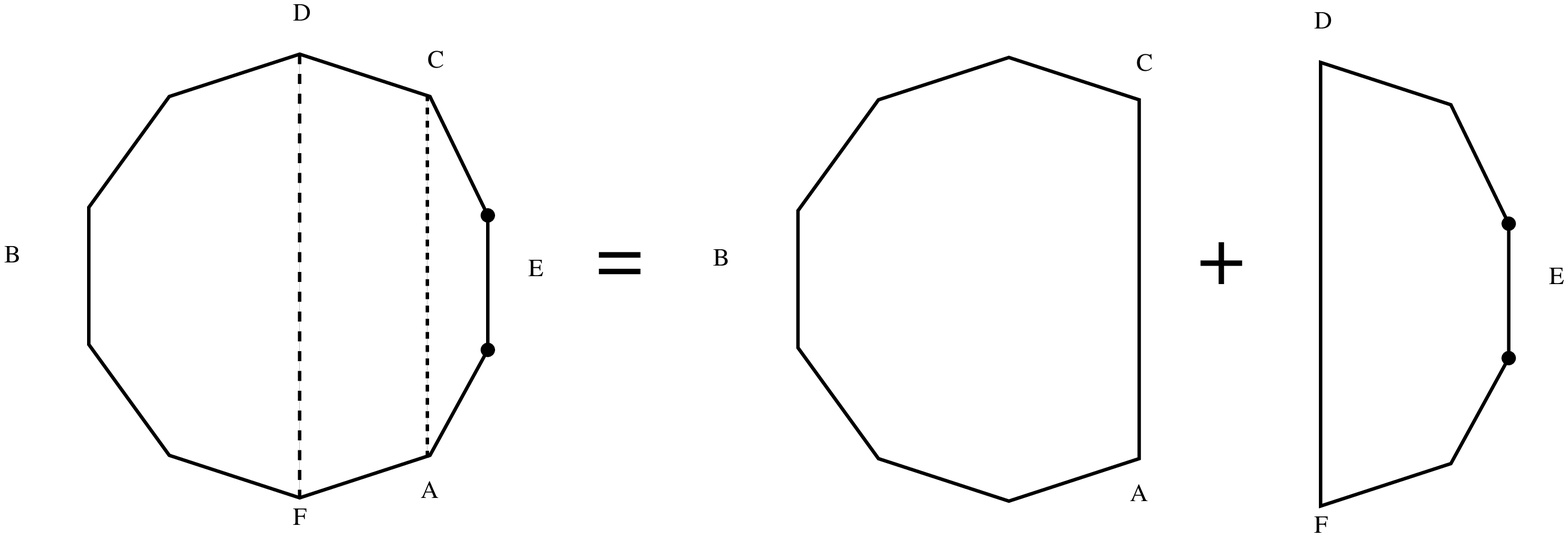}
  \caption{Multi-collinear limit. Here we represent the decomposition
    in~\eqref{coll}. The thick dots represent the $k$
    vertices which approach the dashed
    line AC in the $(k+1)$-collinear limit. }
\label{fig:col}
\end{figure}

One nice property of the remainder function is its very simple transformation properties under collinear limits~\cite{Anastasiou:2009kn}
\begin{equation}\label{coll}
\cR_n\rightarrow\cR_{n-k} + \cR_{k+4}
\end{equation}
 for a $(k+1)$-collinear limit,
ie where $(k+1)$ momenta become collinear. This equation is plotted in
figure~\ref{fig:col}. The first term on the right-hand side is the
reduced $n-k$ polygon emerging in this multi-collinear limit.
Meanwhile the second term arises from the $(k+1)$-collinear splitting
function~\footnote{More precisely the part of the splitting function not
  already contained in the BDS expression.}.
Note that the latter is present because the collinear limit is taken
after expanding the full Wilson loop in $\eps$ (that is the distance
between the
$k$ vertices and the dotted line is limited by the UV cutoff.)

It is a nice feature that
both terms on the right-hand side are themselves remainder functions
of different
ranks. In this multi-collinear limit, the entire set of cross-ratios
of $\cR_n$ decomposes into the set of cross-ratios for $\cR_{n-k}$ and
the set of cross-ratios for $\cR_{k+4}$. These two sets can be most
easily determined from figure~\ref{fig:col}. In particular one draws
all independent $u$-cross-ratios correspnding to the first polygon on
the right-hand side for the first set, and all independent
cross-ratios within the
second polygon for the second set. The first set is independent of the
muti-collinear momenta whereas the second set depends {\em only} on
the multi-collinear variables ($z_1, \dots z_{k+1}$ and the ratios of
vanishing kinematic invariants.)

 For example when $k=2$ we have a
triple collinear limit.
We choose $p_4$, $p_5$ and $p_6$ to be the   collinear momenta, so that
\beq
\label{tripcol567sixpoints}
p_4 := x_4-x_5 = z_1P \  , \quad p_5 = x_5-x_6 = z_2 P \  , \quad p_6 = x_6-x_1 = z_3 P \ , \quad
z_1+z_2+z_3 =1 \ ,
\eeq
and $\cR_{k+4}= \cR_6(\bar{u}_1, \bar{u}_2,
\bar{u}_3)$ where~\cite{seven,Anastasiou:2009kn}
\beq
\label{ubars}
\bar{u}_1 =  {1 \over 1-z_3}{s_{45}  \over s_{456} } \ , \quad
\bar{u}_2 =  {1 \over 1-z_1}{s_{56}  \over s_{456} } \ , \quad
\bar{u}_3 = {z_1z_3 \over (1-z_1)(1-z_3)} \ .
\eeq

A very similar property holds for the ratio of Wilson loops (Amplitudes)
defined here too, with the caveat that one needs to keep both variables, $u$ and $\tilde{u}$ active as
the reference cross-ratios $\tilde{u}$ do change under collinear
limits. Thus we have $w_n (s,u) -w_n^{\rm    ref}(s,\tilde{u}) \rightarrow (w_{n-k} (s,u) -w_{n-k}^{\rm
    ref}(s,\tilde{u}))  + (w_{k+4} (s,\overline u) -w_{k+4}^{\rm
    ref}(s,\overline{\tilde{u}}) )$. For the original and the
  reference Wilson loop to satisfy the same collinear limits, the
  cross-ratios in the splitting functions
  $\overline u$ and $\overline{\tilde{u}}$ must of course be the same
  functions of the multi-collinear variables, $z_1, \dots z_k$
  etc. Thus $\overline u = \overline{\tilde{u}}$ and
\begin{equation}
w_n (s,u) -w_n^{\rm    ref}(s,\tilde{u}) \ \rightarrow \  w_{n-k} (s,u) -w_{n-k}^{\rm
    ref}(s,\tilde{u})\ .
\end{equation}
Thus for the ratio, the collinear limit is purely geometrical with no
contributions from the splitting function as you might expect for a
physical and finite quantity~\footnote{In distinction with the
  standard definition of the remainder function which included
  subtraction of the BDS term expanded in $\epsilon$, here the collinear
  limit can be taken with or without expanding in any UV regulator.}.

\section{A continuous family of regular polygons}
\label{reg}
In general at $n$-points the number of independent cross-ratios,
$n(n-5)/2$, grows quadratically with $n$. It is therefore useful to
identify smaller subsets on this configurations space, guided by
symmetry. One example of such a symmetry is cyclic symmetry,
$Z_n$, which singles out regular polygons.

The authors of~\cite{amsv} argued that for polygons in four
dimensions, there is a one parameter family~\footnote{Allowing
  external particles to lie in an arbitrary number of dimensions gives
higher dimensional families of regular polygons, for example at 8 points we would have a
two parameter family and at 10 points a three parameter family.} of regular polygons for
any even $n$. The family depends continuously on the parameter $\phi$
and varying $\phi$ one covers a particular slice of the $u_{ij}$
space. In particular $\phi=0$ and $\phi=(n-4)\pi/2$ correspond to
special regular polygons, which lie entirely in 3 and 2 dimensional
subspaces respectively.

We will now briefly discuss how one
can introduce general $Z_n$ symmetric kinematics.
To this end we first need to identify a conformal transformation $T$ such
that $T^{\,n}=1$. The vertices of the $Z_n$ symmetric polygon $X^{(p)}$
are obtained by applying $T$ consecutively $p$ times to an arbitrary
point $X^{(0)}$ so that $X^{(p)}=T^{\,p} X^{(0)}$. A natural way to
discuss the action of the conformal group $O(2,4)$ in Minkowski space,
is to describe Minkowski space in terms
of six projective co-ordinates  $X_I$
living in $2+4$
dimensions satisfying
\begin{align}\label{6null}
  X_{-1}^2 + X_0^2 -X_1^2-X_2^2-X_3^2-X_4^2=0\ .
\end{align}
The conformal group $O(2,4)$ then acts linearly on these coordinates
and four-dimensional Minkowski space coordinates can be obtained
straightforwardly from these via~(\ref{x=X}).

By choosing  a suitable basis, a general element of $O(2,4)$ can be
represented in terms of three 2x2 blocks $M_1,M_2,M_3$ which are
either rotations or reflections acting on two coordinates
each.  For
this to satisfy $T^{\,n}=1$ each of the $O(2)$ rotations must be
characterised by an angle $2\pi r/n$ with $r$ an integer and
reflections can only be introduced for $n$ even. For example
a $Z_n$ element composed of two rotations and one reflection acting on
$X$-space is given by
\begin{equation}\label{long}
\left(
\begin{array}{cccccc}
\cos(2\pi r_1/n)&-\sin(2\pi r_1/n)&&&& \\
\sin(2\pi r_1/n)& \cos(2\pi r_1/n)&&&&\\

&&\cos(2\pi r_2/n)&-\sin(2\pi r_2/n)&& \\
&&\sin(2\pi r_2/n)& \cos(2\pi r_2/n)&&\\
&&&&-1&\\
&&&&&1
  \end{array}
\right)
\left(
  \begin{array}{c}
    X_{-1}\\
    X_{0}\\
    X_{1}\\
    X_{2}\\
    X_{3}\\
    X_{4}
  \end{array}
\right)\ .
\end{equation}

The particular configuration of~\cite{amsv} corresponds to the
equation above with $r_1=1$ and $r_2=2$.
In this case with two rotations and one reflection, without loss of
generality we
can choose our initial vertex $X^{(0)}$ as
\begin{align}\label{x0}
  (X^{(0)})^T = (l_3,0,l_2,0,l_1,1)\ .
\end{align}

From now on we will concentrate on this specific $Z_n$-symmetric
family with $r_1=1$ and $r_2=2$
and postpone the Wilson loop computations of other regular
configurations arising from this setup to
future work.

Applying the transformation~\eqref{long} consecutively on~(\ref{x0})
we obtain the polygon vertices
\begin{align}
  X_{-1}^{(p)}&=l_3 \cos \left(\frac{2 \pi  p}{n}\right) &\qquad  X_{0}^{(p)}&=l_3 \sin \left(\frac{2 \pi  p}{n}\right)\nonumber \\
  X_{1}^{(p)}&=l_2 \cos \left(\frac{4 \pi  p}{n}\right) &\qquad  X_{2}^{(p)}&=l_2\sin \left(\frac{4 \pi  p}{n}\right) \label{X=l}\\
  X_{3}^{(p)}&= (-1)^p l_1  &\qquad  X_{4}^{(p)}&=1\nonumber \ .
\end{align}
where $p=1\dots n$ labels the $n$ vertices. Eq~(\ref{6null}) implies that
$l_1,l_2$ and $l_3$ satisfy
\begin{align}\label{l=0}
  1 + l_1^2 +l_2^2 - l_3^2 = 0.
\end{align}
So far the discussion applies to arbitrary regular polygons. We are
specifically interested in null polygons for which $X^{(p)}\cdot
X^{(p+1)}=0$ giving the additional constraint
\begin{align}\label{eq:2}
\qquad
l_1^2 + \sin({2 \pi\over n})^2 l_2^2 - \sin({\pi\over n})^2 l_3^2 = 0 \ .
\end{align}
We have three parameters satisfying two equations, thus our
kinematics depends on one
free parameter. Following the notations of~\cite{amsv} we express $l_1$ in terms of the parameter
$\phi$ as
\begin{align}\label{l1}
  l_1=\tan(\pi/n)\tan(2\pi/n) \tan(\phi/n)\ .
\end{align}
The paramers $l_2$ and $l_3$ are then determined by~(\ref{l=0}) and~(\ref{eq:2}).
Note that the Wilson loop for the $\phi$ family in~\cite{amsv}  was
specified in terms of three complex
variables instead of the six real variables we are using. The
resulting kinematics~(\ref{X=l}),(\ref{l=0}) is
entirely equivalent to the three complex coordinates, but makes more
transparent the
transition between different space-time signatures in four
dimensions. These transitions are necessary when one varies  $\phi$ (or
more generally $u_{ij}$) in the entire parameter space (for example
when $\phi$ exceeds the value $(n-4)\pi/2$ corresponding to the special
two-dimensional regular polygon). In terms of the $l_i$ these
transitions in signature occur when some of the previously real
$l_i$'s become purely imaginary, this leads to the corresponding $X_I$
coordinates becoming purely imaginary, giving the appropriate
sign changes in~(\ref{6null}).

This completes the description of the $\phi$-family geometry in terms
of six-dimensional coordinates.

Below we will  interpret the kinematics
in terms of four dimensional Minkowski space and find the $u_{ij}$
cross-ratios,  in order to compute the corresponding Wilson loops at
weak coupling and compare to the strong coupling results of~\cite{amsv}.

One way to get standard four-dimensional coordinates from these is to
simply divide $X_0\dots X_3$ by $X_{-1}+X_4$ giving
\begin{align}\label{x=X}
  x_\mu={1\over X_{-1}+X_4 }\,X_\mu\ , \qquad  \mu=0\dots 3\ .
\end{align}
One can check that with these definitions that $x^2_{p\,p+1}:=(x^{(p)}-x^{(p+1)})^2=0$ as
required for a null polygon.

Having obtained the vertices of the polygon in four dimensional
Minkowski space we can now find the conformally
invariant cross-ratios
\begin{equation}
  u_{ij}={x_{ij+1}^2 x^2_{i+1 j} \over x_{ij}^2 x_{i+1 j+1}^2}
\end{equation}
in terms of these.

For the regular hexagon, $n=6$ the cross-ratios are
\begin{align}\label{u=phi}
  u_{i i+3} &= {1\over 4} \sec^2(\phi/3) &\qquad i&=1,2,3\ ,
\end{align}
where $\phi$ varies between 0 and $3\pi/2$. The special value
$\phi=\pi$ gives $u=1$ and
corresponds to a regular polygon which can be embedded in $1+1$
dimensions. The point $\phi=0$ gives $u=1/4$ and corresponds to the
special polygon embedded in $1+2$ dimensions. The cross-ratios $u$
become infinite at $\phi=3\pi/2$.

Of course the geometry is defined for any
value of $\phi$ through~\eqref{l1} and one can ask what happens when
$\phi$ exceeds the ``extreme'' point $\phi=3\pi/2$. What happens is
immediately seen from the equation for
$u_{i i+3}$, namely at   $\phi=3\pi/2$,  $u$ reaches infinity, makes a
$u$-turn and
bounces back.

For $n=8$ the cross-ratios are divided into two $Z_n$ invariant
groups and the $\phi$-family is characterised by
\begin{align}
  u_{i i+3} &= {1\over 1+\sqrt 2\cos(\phi/4)} &\qquad i&=1,\dots ,8 \nonumber\\
  u_{i i+4} &= {1 \over 2} &\qquad i&=1,\dots ,4  \ .\label{u8}
\end{align}
Here $\phi$ varies between $0$ (corresponding to a $1+2$ dimensional
polygon) and $3\pi$ where $u_{i i+4}\rightarrow +\infty$. It passes
through the special value $\phi=2\pi$, the special regular polygon in
$1+1$ dimensions.

This time when $\phi$ passes through the ``extreme'' point $3\pi$, the
cross-ratio $u_{i i+3}$ goes from $+\infty$ to $-\infty$ (or
equivalently the cross-ratio $1/u_{i i+3}$ goes through zero.) In this
way the behaviour of the cross-ratios beyond the extreme point is
different from the $n=6$ case considered earlier. There is no
conceptual obstruction to computing the Wilson loop at any values of
$\phi$ but there are practical problems when $u<0$ as this regime can
not be addressed in the fully Euclidean calculation with all two- and
multi-particle invariants negative.

For $n=10$ there are three groups of cross-ratios
\begin{align}
\{u_{ii+3}\}_{i=1}^{10}\ ,   \qquad \{u_{ii+4}\}_{i=1}^{10}\ ,   \qquad
 \{u_{ii+5}\}_{i=1}^{5}\ .
\end{align}
We have not attempted to find simple analytic expressions for the
cross-ratios with $n\geq 10$ but there is no obstacle in doing so
numerically and this will enable us to compute Wilson loops as
a function of $\phi$ at $n=10$ (see section~\ref{10pnt}) and in
principle beyond.
We find that at $n=10$, $\phi$ varies between $0$ and $ 5
{\rm Arccos} (-\sqrt{5}/3)$. At the ``extreme'' point
$\cos(\phi/5)=-\sqrt{5}/3$, $u_{ii+5}$ reaches plus infinity (and
then performs a $u$-turn) while $u_{ii+4}$ goes through zero and
becomes negative.

In the following sections we will compute Wilson loops for hexagonal,
octagonal and decagonal $\phi$-families at one- and two-loops in
perturbation theory and plot them alongside the strong coupling results.

For general $n$-polygons, $\phi$-families start at a special regular polygon
in $(2+1)$ dimensions at $\phi=0$ and as one increases $\phi$ pass through another special polygon
in $(1+1)$ dimensions at $\phi=\pi (n-4)/2$ before they reach the ``extreme'' point where (some of) the $u$-variables
become infinite.
At strong coupling, the ratio of Wilson loops for all $\phi$-families is entirely determined by the free-energy
expression, which is \cite{amsv} a quadratic function of $\phi$
\begin{equation}
  A_{\rm free} \,=\,  -{2\over n \pi}\, \phi^2 \, +\, {\rm const}\ .
\end{equation}

\section{Regular hexagons}
\label{6pnt}
In this section we concentrate on the $\phi$-family of regular
hexagons. In a later part of the section we will present the results
in terms of the $\phi$ parameter, but first we want to comment on the
existing results in the literature which are naturally expressed in
terms of the standard cross-ratios $u$. The $\phi$-family corresponds
to all $u$'s equal and varying between $u=1/4$ and $u=\infty$.
By extending $\phi$ to have imaginary values one can cover the full
range from $u=0$ to $u=\infty$.

At  weak coupling the remainder
function starts at two-loops. It was first computed
numerically in~\cite{dhks6} where it was found to
agree with the MHV amplitude computed in~\cite{seven}.
In~\cite{Anastasiou:2009kn} a formalism was developed for the
numerical computation of general $n$-gon Wilson loops at the two loop
level. Detailed plots of general $u$'s are known and in particular for
all $u$'s equal we can cover the entire interval
$0<u<\infty$. Furthermore analytical results have now been derived for
$0<u_1, u_2, u_3<1$ and also in certain limits where $u_i\rightarrow
\infty$~\cite{6an1,6an2}. The combined results of~\cite{Anastasiou:2009kn}
and~\cite{6an2} are plotted below in figure~\ref{fig1}.

The strong
coupling remainder function $\cR_6(u_1,u_2,u_3)$ has also been derived
in~\cite{agm} using integrable techniques. For all $u$'s equal
it
takes a surprisingly simple form
\begin{align}\label{rstrong}
  \cR_6^{\mathrm{strong}}(u,u,u)= {\pi \over 6} - {1 \over 3 \pi}
  \phi^2 -{3\over 8} (\log^2(u) + 2Li_2(u)) + \mathrm{const}
\end{align}
where $\phi(u)$ is defined in~(\ref{u=phi}).

It is interesting to
compare the dependence of the strong and weak coupling results on the
kinematics. The authors of~\cite{agm} made a very interesting
observation.
By modifying the strong coupling results~(\ref{rstrong}) by introducing
three coefficients $c_1,c_2$ and $c_3$,
\begin{align}\label{rstrongc}
  \cR_6^{\mathrm{AGM}}(u,u,u)= c_1(-{\pi \over 6} +{1 \over 3 \pi}
  \phi^2) + c_2({3\over 8} (\log^2(u) + 2Li_2(u))) + c_3\ ,
\end{align}
a very close match with the weak coupling remainder can be found
for particular chosen values for $c_i$.

The constant $c_3$ is fixed by the collinear limit $\cR_6 \rightarrow
\cR_5$ and with a little work\footnote{The collinear limit brings one
  outside the all $u$ equal regime.} can be found to be $c_3=-c_2 \pi^2/12$.
We plot the combined weak coupling result and the AGM expression
for $c_1=0.263 \pi^3$ and $c_2=0.860 \pi^2$.

\begin{figure}[h]
  \centering
  \includegraphics[width=10cm]{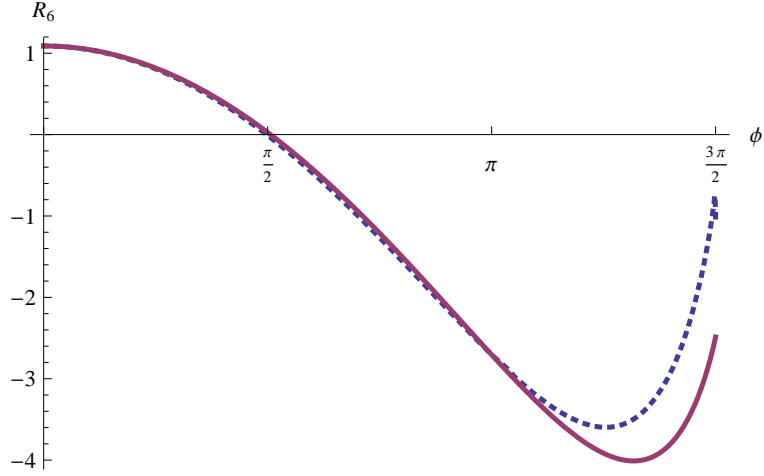}
  \caption{The remainder function for the $\phi$-family at six
    points. The dashed line gives the two-loop result and the solid
    line is the AGM modified strong coupling expression. }
\label{fig1}
\end{figure}

The modified strong coupling result~(\ref{rstrongc}) and the weak
coupling result at two loops, $\cR^{(2)}$ (recall there are no one loop
contributions to the remainder) are close to each other but
unfortunately can never be made identical by any choice of
coefficients. The strong coupling result~(\ref{rstrong}) is not of a
uniform transcendental weight. However each separate term in the modified AGM
expression~(\ref{rstrongc}) has a uniform transcendental weight which
can be made equal to four by choosing $c_1\sim \pi^3$ and $c_2\sim
\pi^2$. It is important to note that there are no genuine weight four functions (eg no
$\Li_4$ or $\log \times \Li_3$) in this
expression. This is to be contrasted with the analytic form of
the weak coupling result~\cite{6an1,6an2} which is of transcendental weight
four and
where functions with intrinsic
weight four do appear. For example the value of $\cR^{(2)}$ for the
special regular polygon at $\phi=0$ (ie $u=1/4$) is~\cite{6an2}
{\small \begin{align}
 \cR_{6}^{(2)}&\left({1\over 4},{1\over 4},{1\over 4}\right) =3 \text{Li}_2\left(\frac{1}{3}\right) \log ^2 2 -\frac{9}{2} \text{Li}_2\left(\frac{1}{3}\right) \log ^2 3 -\frac{567}{4} \text{Li}_3\left(\frac{1}{3}\right) \log  2\nonumber\\
& +\frac{543}{4} \text{Li}_3\left(-\frac{1}{2}\right) \log  2 +\frac{567}{8} \text{Li}_3\left(\frac{1}{3}\right) \log  3 -\frac{567}{4} \text{Li}_3\left(-\frac{1}{2}\right) \log  3 +\frac{1323}{16} \zeta_3 \log  2\nonumber\\
& +\frac{945}{32} \zeta_3 \log  3 -\frac{39}{32} \log ^4 2 -\frac{257}{64} \log ^4 3 +\frac{173}{8} \log  3  \log ^3 2 +\frac{189}{8} \log ^3 3  \log  2 -\frac{543}{16} \log ^2 3  \log ^2 2\nonumber\\
& -\frac{63}{16} \pi ^2 \log ^2 2 -\frac{181}{64} \pi ^2 \log ^2 3 +\frac{189}{2} \text{Li}_4\left(\frac{1}{2}\right)+\frac{1701}{8} \text{Li}_4\left(\frac{1}{3}\right)-\frac{543}{16} \text{Li}_4\left(-\frac{1}{3}\right)\nonumber\\
&+\frac{555}{2} \text{Li}_4\left(-\frac{1}{2}\right)-\frac{9}{2}
\text{Li}_2\left(\frac{1}{3}\right)^2-\frac{567}{16}
S_{2,2}\left(-\frac{1}{3}\right)-\frac{567}{4}
S_{2,2}\left(-\frac{1}{2}\right)-\frac{2123 \pi ^4}{2880}\ .
\end{align}
}
Interestingly the value for the other special regular polygon at
$\phi=\pi$ (ie $u=1$)
is much simpler~\cite{Anastasiou:2009kn,6an2}
\begin{align}
\cR_{6}^{(2)}(1,1,1)=-\pi^4/36\ .
\end{align}

\begin{figure}[t]
  \centering
  \includegraphics[width=10cm]{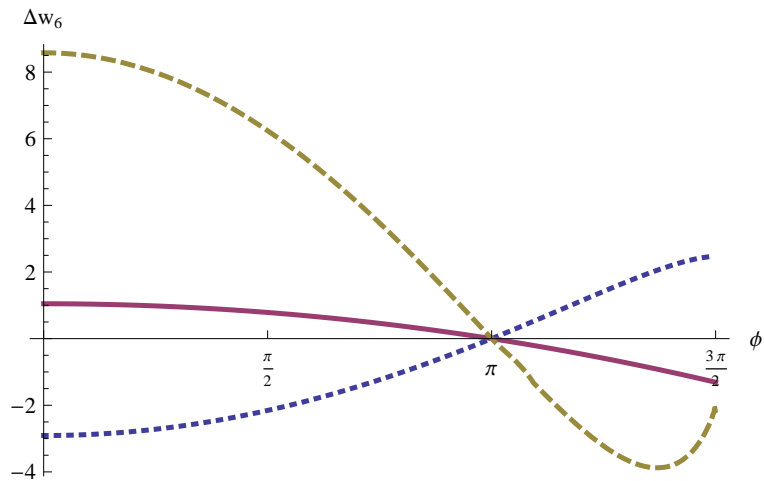}
  \caption{The Wilson loop ratio for the $\phi$-family at six
    points. This graph shows
    $\Delta w_6$ at one loop (dotted), two loops (dashed)  and at
    strong coupling (solid line). The two special regular polygons are
    at $\phi=0$ and at $\phi=\pi$, the latter being chosen as the
    reference point.}
\label{fig:dw6}
\end{figure}

What was the rationale for introducing different coefficients in the
strong coupling result? The two terms in the remainder function have a
completely different origin, the first being the free energy of an
appropriate integrable system, while the second arose from subtracting
the BDS expression from the cutoff area. As already discussed in
section~\ref{sec:1} the BDS expression is essentially dictated by the
one-loop result and is not the object that appears naturally at strong
coupling. One may hope that there is a meaning in expressing
both the strong and the weak coupling results as a linear
combination of certain master functions and then fitting the
coefficients. But it is clear that the simple division involving $c_1$
and $c_2$ described above does not work. At present we lack the theory
giving the basis of master functions (at least at weak coupling).

We will thus concentrate on the entire Wilson loop (regularised by
dividing by  the reference Wilson loop as explained in section~\ref{sec:2}). Whilst the
BDS remainder function was not a natural object at strong coupling,
the Wilson loop ratio is natural at both strong and weak coupling.
In fact at strong coupling, for the regular Wilson loops we are considering, the  log of the Wilson
loop ratio is just the free energy
\begin{align}\label{DW6}
  \log \left({W_6 \over W_6^{\rm ref}}\right)\  :=\  w_6 -w_6^{\rm
    ref}\ = \ A_{\rm free} - \mathrm{
    const} \ = \  - {1 \over 3 \pi}
  \phi^2 + {\pi \over 3}\ .
\end{align}
Here we have treated the cross-ratios for the reference Wilson loop
$\tilde{u}$ as fixed. For all $\phi$-families throughout the
paper we will choose all the reference Wilson loops to be the Wilson
loop of the special regular polygon in $1+1$ dimensions~\footnote{This
choice determines the constant on the right-hand side of (\ref{DW6})}, in other
words $\phi_{\rm ref} = (n-4)\pi/2$. At strong coupling there is only
the free energy left and thus no clear reason for introducing more
than one coefficient.

The Wilson loop ratio is also a natural object to compute at weak
coupling. Indeed, in all computations of the remainder
function, it is the Wilson loop which is computed directly and not
the remainder function.
The Wilson loop ratio has a non-trivial contribution already at one
loop
\begin{align}\label{Dw61}
  w_6^{(1)} - w_6^{(1) \mathrm{ref}}  =
    -{\gamma_K^{(1)}\over 2} \left\{ {3\over 8} \left[\log^2(u) + 2Li_2(1-u)\right]
    \right\} \ .
\end{align}
This arises from the BDS expression (see Appendix).

The two loop
expression $\Delta w_6^{(2)}(\phi):= w_6^{(2)} - w_6^{(2) \mathrm{ref}}$
we now compute numerically as in~\cite{Anastasiou:2009kn}. One
can always extract the remainder function from $\Delta w_6^{(2)}(\phi)$ by
subtracting~(\ref{Dw61}) with  $\gamma_K^{(1)}$ replaced by $\gamma_K^{(2)}$.

Figure~\ref{fig:dw6} gives results for $\Delta w_6^{(2)}(\phi)$ at one-loop,
two-loops and at strong coupling. There are no coefficients, the three
curves are distinct but it is interesting that the two loop result
(whose analytic form consists of about 100 terms
involving multiple
polylogarithms) has a deceptively simple looking graph. Indeed between
the two special regular polygons at $\phi=0$ and $\phi=\pi$ the
function can be well approximated by a quadratic function just as at
strong coupling~(\ref{DW6}). Furthermore, in the interval between
the two special regular polygons all three contributions are almost
identical up to rescaling.

We now proceed with the perturbative analysis of the $\phi$-families\footnote{In~\cite{amsv} the
  $\phi$-family was actually only defined within the
  interval between the two special regular polygons, while we continue increasing $\phi$ beyond the
  ``extreme'' point.} at $n=8$ and $n=10$.

\section{Regular octagons}
\label{8pnt}

\begin{figure}[h]
  \centering
  \includegraphics[width=5cm]{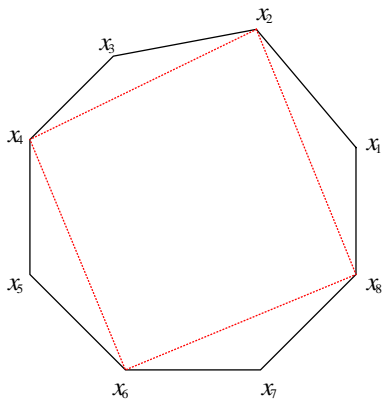}
  \caption{The cross-ratio $\chi^+/\chi^-$ is a rectangle with all edges made out of two-particle invariants.
  The second two-particle invariant rectangle, $x_{35}^2 x_{17}^2 / (x_{13}^2 x_{57}^2),$ is obtained by the
  cyclic relabelling $x_i \to x_{i+1}$ and corresponds to the combination  $\chi^+  \chi^-.$}
\label{fig:8pntchi}
\end{figure}

The $n=8$ case is the simplest example of  a polygon with $n$
divisible by four, which is a special case since the entire kinematics
can not be cleanly separated into $s$ and $u$ variables. In particular
for the octagon case at hand we have two cross-ratios, $\chi^+$ and
$\chi^-$ which are made entirely from two-particle invariants. In the case of the octagon
one can think of two rectangles with edges made out of two-particle invariants only, which
can be embedded into the octagon. Figure~\ref{fig:8pntchi} shows one of these rectangles,
and the second one can be obtained by a cyclic relabelling $x_i \to x_{i+1}$ of all vertices on that figure.
In terms of the variables $\chi^\pm$ used in \cite{am8,bhkt}, the two rectangular cross-ratios are
$\chi^+  \chi^-$ and $\chi^+ / \chi^-$, 
\begin{align}
  \chi^+ \chi^- = {x_{35}^2 x_{17}^2 \over x_{13}^2 x_{57}^2} \ = \ {
    u_{15} u_{25} u_{16} u_{26}  \over u_{37} u_{38} u_{47}
    u_{48} }
\qquad \qquad
  {\chi^+ \over \chi^-} = {x_{24}^2 x_{68}^2 \over x_{46}^2 x_{28}^2} \ = \ {
    u_{48}  u_{58} u_{14} u_{15}  \over u_{26} u_{27} u_{36} u_{37} }
  \ .
\end{align}
The combination $\chi^+ / \chi^-$
is depicted in figure~\ref{fig:8pntchi}.

In general we can not vary these two cross-ratios independently of
$s$. However in the regular octagon case~(\ref{u8}) these cross-ratios
are fixed, $\chi^+=\chi^-=1$. Instead the cross-ratios we vary when we
vary $\phi$ are $u_{i i+3}$ and can be varied  independently of $s$.

We now consider the  log of the ratio of the
Wilson loops
\begin{align}
  \Delta w_8(\phi) := w_8(s;\phi)-w_8 (s,\phi^{\rm{ref}}=2 \pi)\ ,
\end{align}
where as always the reference point was chosen to be the $1+1$
dimensional special polygon.
The strong coupling result for $\Delta w$ is again simply the free
energy of~\cite{amsv} (all other contributions cancel)
\begin{align}
  \Delta w_8^{\mathrm{strong}}= A_{\mathrm free} -{\rm{const}}=\pi
  -\frac{\phi ^2}{4 \pi }\ .
\end{align}

\begin{figure}[h]
  \centering
  \includegraphics[width=10cm]{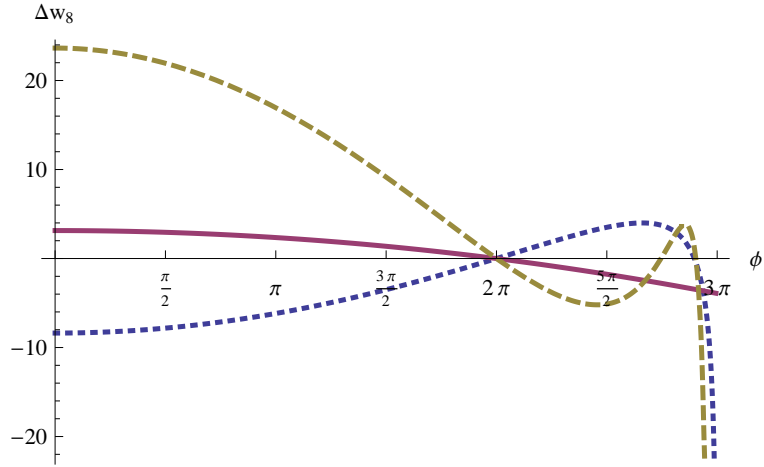}
  \caption{The Wilson loop ratio for the $\phi$-family at eight
    points. This graph shows
    $\Delta w_8$ at one loop (dotted), two loops (dashed)  and at
    strong coupling (solid line). The two special regular polygons are
    at $\phi=0$ and at $\phi=2\pi$, the latter being chosen as the
    reference point.}
\label{fig:dw8}
\end{figure}

At weak coupling we have computed the one- and two-loop contributions.
The one loop result is
\begin{align}
\Delta w_8^{(1)}=  -\gamma_K^{(1)} \left(\text{Li}_2(1-u_{14})+\log
  \left(\frac{u_{14}}{2}\right) \log (u_{14})\right)\ , \qquad
u_{14}={1 \over 1+\sqrt 2 \cos (\phi/4)}
\end{align}
and the two loop result is computed numerically. All three results are
displayed in figure~\ref{fig:dw8}. Once again all three curves can be well
approximated by a quadratic in the interval between the two special
polygons. However, as $\phi$ reaches the extreme value the weak
coupling and strong coupling results diverge.

Before we conclude this section we would like to comment on a
different type of eight point kinematics, previously considered
in~\cite{am8,bhkt} at strong and weak coupling correspondingly. This
kinematics corresponds to polygons which can be embedded in the
boundary of $AdS_3$. In contrast  to the $\phi$-kinematics discussed
above, here $\chi^+$ and $\chi^-$ vary while all other cross-ratios
are fixed, see section 3.1 of~\cite{bhkt} for details. This implies
that in this kinematics, the only free cross-ratios, $\chi^+$ and
$\chi^-$ are the ones which are entirely determined by the
two-particle invariants, hence the ratio of Wilson loops we are
considering here is trivial and thus trivially agrees between strong
and weak coupling. It is curious that in this special kinematics the
two-loop remainder numerically matched the strong coupling
counter-part up to an overall rescaling~\cite{bhkt}. It is tempting to
compare this situation to results for $n<6$ where the ratio is also
trivial as there are no multi-particle invariants. These are the cases
where strong and weak coupling results also agree as they are given by
the BDS expression.

\section{Regular decagons}
\label{10pnt}

\begin{figure}[h]
  \centering
  \includegraphics[width=10cm]{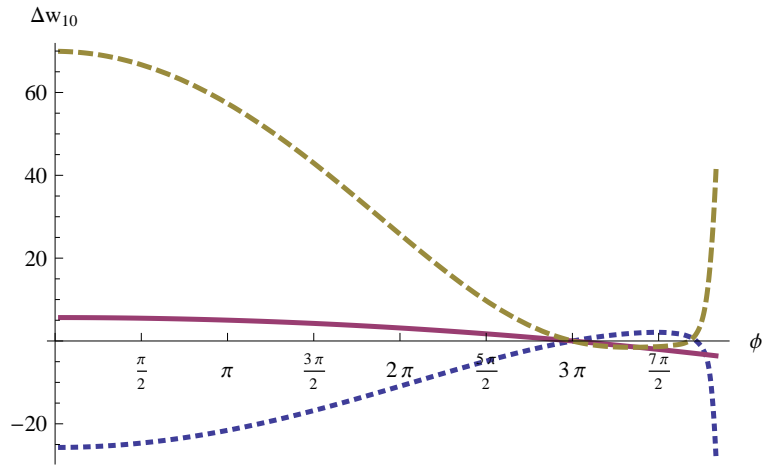}
  \caption{The Wilson loop ratio for the $\phi$-family at eight
    points. This graph shows
    $\Delta w_{10}$ at one loop (dotted), two loops (dashed)  and at
    strong coupling (solid line). The two special regular polygons are
    at $\phi=0$ and at $\phi=3\pi$, the latter being chosen as the
    reference point.}
\label{fig:dw10}
\end{figure}

The final set of computations which we have performed for the $\phi$-family
is for $n=10$ regular polygons and our strategy of dividing by the
reference area works without exceptions. The strong coupling result
for $\Delta w_{10}$ is once again determined by the free energy of
reference~\cite{amsv} and is quadratic in $\phi$. The one loop result
can be determined from the BDS expression, and the two-loop result we
have computed. These three contributions are plotted alongside each
other in figure~\ref{fig:dw10}. As always, the one-loop result should be multiplied
by $a$, the two-loop result by $a^2$ and the strong coupling result by
$\sqrt{2a}$.

Once again between the two special polygons at $\phi=0$ and
$\phi=3\pi$ there is a similarity between the three curves, though
the two loop curve is starting to visibly deviate from the quadratic
form. Beyond $3\pi$ and especially as $\phi$ approaches the extreme
point, the weak coupling results start to diverge, whilst the strong
coupling result does not.

To summarise, in this paper we have proposed a new finite,
regularisation independent and  conformally
invariant way to characterise the Wilson loops and amplitudes in both
weak and strong coupling regimes. As an application, we have computed
Wilson loops for regular polygons, up to $n=10$ and have compared our
weak coupling results to the recently derived strong coupling
expressions of~\cite{amsv}.

\vspace{.3cm}

\section*{Acknowledgements}

It is a pleasure to thank  Babis Anastasiou, Andreas Brandhuber,
Patrick Dorey,
Claude Duhr,
Bill Spence and Gabriele Travaglini
for discussions at various stages of this work.
VVK acknowledges a Leverhulme Research Fellowship.

\newpage

\section*{Appendix}

\appendix

\section{BDS}

For the convenience of the reader we here define the BDS expressions
appearing in section~\ref{sec:1}. In the context of MHV amplitudes
in dimensional regularisation with $D=4-2\epsilon$ we have~\cite{bds}
\begin{align}\label{a1}
  (BDS)_n \ =  \
 \sum_{L=1}^{\infty} a^L \, f^{(L)} (\epsilon) \, \cM_{n}^{(1)} ( L
 \epsilon) \ + \ C(a)\ .
\end{align}
Here $\cM_{n}^{(1)}(\epsilon)$ is the one loop amplitude,
$a$ is the coupling constant
$ a=[{g^2 N/ (8 \pi^2)}] (4\pi e^{-\gamma})^\eps$ and
\beq \label{fleps}
f^{(L)}(\epsilon ) \, :=\, f_0^{(L)} + f_1^{(L)} \epsilon + f_2^{(L)} \epsilon^2  \ .
\eeq
The $f_i^{(L)}$ are numbers,
in particular, $f_0^{(L)} = \gamma_{K}^{(L)} / 4$, where $\gamma_{K}$ is the cusp anomalous dimension,
\beq
\label{defgamK}
\gamma_{K} (a) =\, \sum_{L=1}^{\infty} \, a^L\, \gamma_{K}^{(L)}\ , \qquad
\gamma_{K}^{(1)}=4 \ , \quad \gamma_{K}^{(2)}= -4\,  \zeta_2 \ ,
\eeq
and $C(a)$ is a (coupling dependent) constant starting from two loops, $C^{(2)}=-\zeta_2^2/2$.

The one-loop amplitude contains an infrared divergent part and a
finite part
\begin{align}\label{m1}
\cM_n^{(1)}(\epsilon)&= -{1\over 2 \e^2}\sum_{i=1}^n   \left( - {s_i}
    \over \mu_\mathrm{}^2 \right)^{-\e} +\ F_n^{(1)}(\epsilon)\ ,\\
F_n^{(1)}(0)& = {1 \over 2} \sum_{i=1}^n g_{n,i} \ ,
\end{align}
where
\beq
g_{n,i}  \ = \
-\sum_{r=2}^{[ n/2 ]  -1}
  \ln \left( { -t^{[r]}_{i}\over -t^{[r+1]}_{i} }\right)
  \ln \left({ -t^{[r]}_{i+1}\over -t^{[r+1]}_{i} }\right) \, + \,
D_{n,i} \, + \, L_{n,i} + {3\over 2} \zeta_2 \  ,
\eeq
and  $t^{[r]}_{i} := (p_i + \cdots + p_{i+r-1})^2$
are the kinematical invariants, $t^{[2]}_i=s_i$ and $t^{[r>2]}_i$ are
multi-particle invariants.  The functions $D_{n,i}$ and $L_{n,i}$
for even values of $n$ are~\cite{bddk}
\beqa
\label{LDeven}
D_{2m,i} &=& -\sum_{r=2}^{m-2}
\Li \left( 1- { t^{[r]}_{i} t^{[r+2]}_{i-1}
\over t^{[r+1]}_{i} t^{[r+1]}_{i-1} }  \right)
- {1 \over 2} \Li \left( 1- { t^{[m-1]}_{i} t^{[m+1]}_{i-1}
\over t^{[m]}_{i} t^{[m]}_{i-1}} \right) \ ,
\\ \nonumber
L_{2m,i} &=& - {1\over 4}
  \ln \left({ -t^{[m]}_{i}\over -t^{[m]}_{i+m+1}  } \right)
  \ln \left({ -t^{[m]}_{i+1}\over -t^{[m]}_{i+m} } \right) \ .
\eeqa
It is clear from~(\ref{a1},\ref{m1})
that to all orders in perturbation theory the divergent
part of the BDS expression depends on the kinematics only through
two-particle invariants, and the finite part is given by
\beq
\label{fbds}
F^\mathrm{BDS}_n (a) = \, {1\over 4} \gamma_K(a)\, \, F^{(1)}_n (0)\
+\ C(a)
\ .
\eeq

The BDS expression for the Wilson loop, $(BDS)_n^{WL}$  has the same
form as
that of the amplitude, $(BDS)_n$ except that the one loop
amplitude $\cM_n^{(1)}$ is substituted by the one loop expression for
the Wilson loop $W_n^{(1)}$ and the coefficient functions
$f(\epsilon)$ are different, $f_{WL}(\epsilon)$,
\begin{align}
  f^{(1)}(\epsilon) & = 1  & \qquad \qquad f^{(2)}(\epsilon)&=-
  \zeta_2 -\zeta_3 \epsilon -\zeta_4 \epsilon^2\nonumber \\
 f_{WL}^{(1)}(\epsilon) & = 1  & \qquad \qquad f_{WL}^{(2)}(\epsilon)&=-
  \zeta_2 +7\zeta_3 \epsilon -5\zeta_4 \epsilon^2\ .
\end{align}
The one loop contribution to the Wilson loop agrees with the one loop
amplitude up to a constant~\cite{dks} for any $n$~\cite{bht}
\begin{align}
  W_n^{(1)}=\cM_n^{(1)} - n {\pi^2 \over 12}\ .
\end{align}
This statement is of course the one loop manifestation of the Wilson
loop/amplitude duality. At higher loops as is by now well known, the
BDS expressions for both the amplitude and the Wilson loop need to be
extended by introducing remainder functions
as in~(\ref{eq:1}). The statement of the duality beyond one loop can
be characterised by the equality of the remainder functions (see
equation~(\ref{r=r}).

\end{document}